\documentclass[conference]{IEEEtran}
\IEEEoverridecommandlockouts
\usepackage{cite}
\usepackage{lineno} 
\usepackage{amsmath,amssymb,amsfonts}
\usepackage{ntheorem}
\usepackage{algorithmic}
\usepackage{graphicx}
\usepackage{textcomp}
\usepackage{xcolor}
\usepackage[ruled,linesnumbered]{algorithm2e}
\usepackage{marvosym}
\usepackage{epstopdf}
\newtheorem{definition}{Definition}
\newtheorem{theorem}{Theorem}
\newtheorem{lemma}{Lemma}

\newtheorem*{proof}{Proof}

\def\BibTeX{{\rm B\kern-.05em{\sc i\kern-.025em b}\kern-.08em
    T\kern-.1667em\lower.7ex\hbox{E}\kern-.125emX}}
\begin{document}


\title{Community Preserved Social Graph Publishing with Node Differential Privacy}

\author{\IEEEauthorblockN{Sen Zhang, Weiwei Ni\textsuperscript{\Letter}, Nan Fu}
\IEEEauthorblockA{\textit{School of Computer Science and Engineering, Southeast University, Nanjing 211189, China} \\
Email: \{senzhang, wni, funan1988\}@seu.edu.cn}
}

\maketitle

\begin{abstract}
The goal of privacy-preserving social graph publishing is to protect individual privacy while preserving data utility. Community structure, which is an important global pattern of nodes, is a crucial data utility as it serves as fundamental operations for many graph analysis tasks. Yet, most existing methods with differential privacy (DP) commonly fall in edge-DP to sacrifice security in exchange for utility. Moreover, they reconstruct graphs from the local feature-extraction of nodes, resulting in poor community preservation.
Motivated by this, we propose PrivCom, a strict node-DP graph publishing algorithm to maximize the utility on the community structure while maintaining a higher level of privacy. Specifically, to reduce the huge sensitivity, we devise a Katz index-based private graph feature extraction method, which can capture global graph structure features while greatly reducing the global sensitivity via a sensitivity regulation strategy. Yet, with a fixed sensitivity, the feature captured by Katz index, which is presented in matrix form, requires privacy budget splits. As a result, plenty of noise is injected, thereby mitigating global structural utility.
To this end, we design a private Oja algorithm approximating eigen-decomposition, which yields the noisy Katz matrix via privately estimating eigenvectors and eigenvalues from extracted low-dimensional vectors. Experimental results confirm our theoretical findings and the efficacy of PrivCom.
\end{abstract}


\section{Introduction}
With the advancement of information technologies, various social graphs are prevalent in many real applications (e.g., social media and knowledge bases). An important property of these graphs is the community. Essentially, a community is a group of nodes, within which nodes are densely connected yet between which they are linked sparsely. In real-world social graphs, nodes in the same community typically share common properties or play similar roles. Hence, it is worth pointing out that the community structure is an important global pattern of nodes \cite{tu2019a}, which is expected to benefit graph analysis tasks, such as community detection for event organization and friend recommendation. While the potential benefits of community are tremendous, it, unfortunately, has been shown that, with naively sanitized graph data, an adversary is able to launch different types of privacy attacks that re-identify nodes, reveal edges between nodes \cite{hay2011privacy}. Thus, before being released to the public, graph data needs to be sanitized with formal, provable privacy guarantees. \\
\hspace*{1em}Differential privacy (DP) \cite{dwork2014algorithmic} is a widely accepted privacy model, which ensures that the output of the process undergoes sufficient perturbation to mask the presence or absence of any individual in the input while offering both provable utilities on the results and practical algorithms to provide them. When applying DP to graph data, two variants of DP are introduced \cite{hay2009accurate}:
in edge-DP, two graphs are neighboring if they differ on a single edge;
in node-DP, two graphs are neighboring if they differ up to all edges connected to one node. Obviously, satisfying node-DP is tremendously difficult,
as removing one node may cause, in the worst case, the removal of $\left|\mathcal{V}\right|-1$ edges, where $\left|\mathcal{V}\right|$ is the set of all nodes.
In comparison, edge-DP is easy to implement and has a good balance with utility preservation. Due to this reason, most existing approaches that publish graphs consider edge-DP. However, the privacy offered by edge-DP is much weaker than node-DP, and is insufficient in most settings. Particularly, edge-DP offers effective protection only when the edges are independent from each other \cite{li2013membership}, whereas it is hard to make the case that edges are independent. In contrast, as demonstrated in \cite{day2016publishing}, node-DP can provide better privacy protection. The main reason is that a node and its associated edges represent all personal data of an individual, and an edge does not represent all data controlled by one individual.
Furthermore, most existing methods reconstruct graphs from the local feature-extraction of nodes instead of global structural features, leading to poor community preservation.
For example, Sala et al. \cite{sala2011sharing} and Wang et al. \cite{wang2013preserving} employ the dK-2 series feature-extraction model to transform the social graph into pairs of node degrees. The dK-2 model keeps the information of the degree; then the published graph has a similar degree distribution. However, the dK-2 model only contains the linking information of two nodes, and contains little to the global pattern of nodes.
Xiao et al. \cite{xiao2014differentially} propose a similar model with a hierarchical random graph (HRG) model to capture the probability of connectivity of nodes. The HRG model stores a cluster of nodes in the same branch on the HRG tree. Unfortunately, the extracted cluster cannot effectively preserve the community structure. The main reason of information loss is the low probability of each single HRG, so that the inferred graph has low probability to be similar to the original one. \\
\hspace*{1em}Inspired by this, we propose PrivCom, a graph publishing algorithm with strict node-DP to maximize the utility on the community structure while guaranteeing a higher level of privacy. More concretely, to reduce the huge sensitivity, we devise a Katz index-based private graph feature extraction method, which can capture global graph structure features while greatly reducing the global sensitivity through a sensitivity regulation strategy. Nevertheless, with a fixed sensitivity, the feature captured by Katz index, which is presented in the form of a matrix, requires privacy budget splits. As a result, a prohibitive amount of noise is added, therefore mitigating global structural preservation.
To fill this gap, we design a private Oja algorithm approximating eigen-decomposition, which generates the noisy Katz matrix via privately estimating eigenvectors and eigenvalues from extracted low-dimensional vectors.
Through formal privacy analysis, we prove that PrivCom satisfies $\left(\varepsilon,\delta\right)$-node-DP. Extensive experimental results confirm our theoretical findings and the efficacy of PrivCom. Briefly speaking, we make the following contributions:
\begin{itemize}
\item[$\bullet$] We present PrivCom, a strict node-DP graph publishing algorithm, which can preserve high data utility on community structure while satisfying $(\varepsilon,\delta)$-DP.
\item[$\bullet$] To reduce the huge sensitivity, we devise a Katz index-based private graph feature extraction method, which can capture global graph structure features while greatly reducing the global sensitivity via a sensitivity regulation strategy.
\item[$\bullet$] With a fixed sensitivity, the generated Katz matrix requires privacy budget splits and thus mitigates global structural utility. For this, we design a private Oja method approximating eigen-decomposition, which yields the noisy Katz matrix via privately estimating eigenvectors and eigenvalues from extracted low-dimensional vectors.
\end{itemize}
\section{Related work}\label{Related_work}
Some latest studies attempt to address the graph synthesis publishing with DP. For instance, Sala et al. \cite{sala2011sharing} propose an alternative approach that makes use of the dK-2 graph model. Wang et al. \cite{wang2013preserving} further improve the work of Sala et al. by considering global sensitivity instead of local sensitivity. In particular, the dK-2 model keeps the information of the degree; then the published graph has a similar degree distribution. Nevertheless, the dK-2 model only contains the linking information of two nodes, and contains little to the global pattern of nodes, which leads to poor community structure preservation.
Xiao et al. \cite{xiao2014differentially} propose a similar model with a hierarchical random graph (HRG) model to capture the probability of connectivity of nodes. The HRG model stores a cluster of nodes in the same branch on the HRG tree. Unfortunately, the extracted cluster cannot effectively preserve the community structure. The main reason of information loss is the low probability of each single HRG, so that the inferred graph has low probability to be similar to the original one.
In addition, this approach above, as well as the subsequent line of work on private graph publishing \cite{wang2013differential}, \cite{chen2014correlated}, only guarantees the weak edge-DP, since DP offers effective protection only when the edges are independent from each other \cite{li2013membership} yet it is hard to make the case that edges are independent. \\
\hspace*{1em}To summarize, how to preserve high utility on the community of graphs and protect individuals with node-DP simultaneously in published graphs remains open in the literature.
\section{Problem Statement}\label{Pro_sta}
\begin{figure}[h]
  \centering
  \includegraphics[width=3.2in]{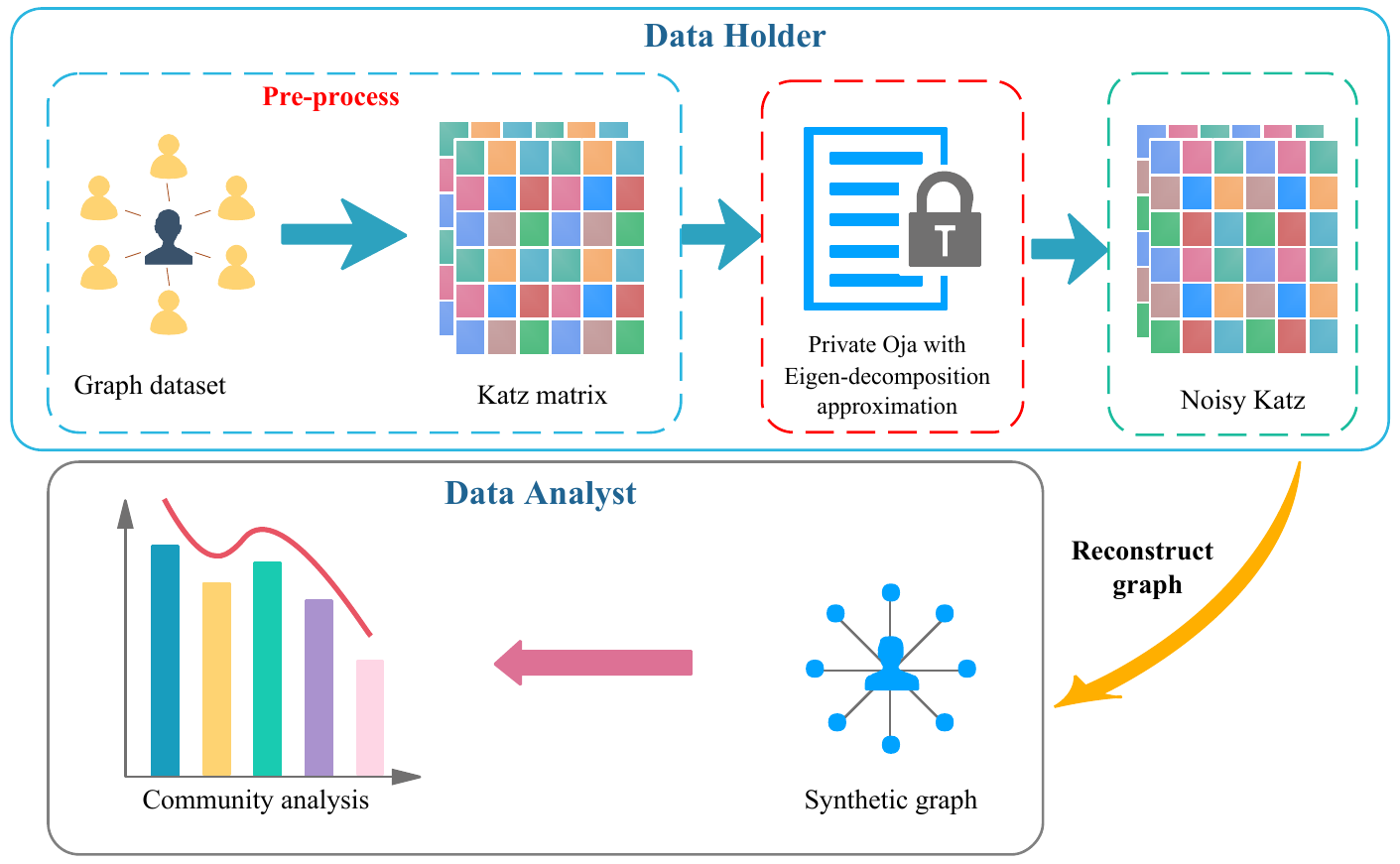}
  \caption{The Framework of PrivCom.} \label{PrivCom_fig}
\end{figure}

In this section, we formally describe our problem setting. Given an undirected graph $\mathcal{G}=\left(\mathcal{V},\mathcal{E}\right)$, we want to release a sanitized graph $\widetilde{\mathcal{G}}$ that approximates the true community of the original graph $\mathcal{G}$ as closely as possible while satisfying the strict node-DP.
Following a priori conclusion that the community structure is an important global pattern of nodes \cite{tu2019a},
we devise a Katz index-based private graph feature extraction method, which can capture global graph structure features while greatly reducing the global sensitivity via a sensitivity regulation strategy.
Then, with a fixed sensitivity, the feature matrix generated by Katz index, requiring privacy budget splits, potentially mitigates global structural utility. For this, we design a private Oja algorithm approximating eigen-decomposition, which yields the noisy Katz matrix via privately estimating eigenvectors and eigenvalues from extracted low-dimensional vectors.
Finally, based on the noisy Katz matrix, we reconstruct a synthetic graph. The process of privately releasing social graphs is illustrated in Fig. \ref{PrivCom_fig}. After reaping processed graph data, the analyst can perform graph analysis tasks, including but not limited to graph community detection.
\section{Preliminaries}\label{Preliminary}
\subsection{Notation}\label{Notat}
Let $\mathcal{G=\left(V,E\right)}$ denote a simple undirected graph, where $\mathcal{V} = \left\{v_1,v_2,\ldots,v_n \right\}$ is the set of nodes and $\mathcal{E \subseteq V \times V}$ is the set of edges. Let $\mathbf{A}\in\left\{0,1\right\}^{n \times n}$ denote the adjacency matrix of the graph $\mathcal{G}$, where $\mathbf{A}_{i,j}=1$ if there exists an edge between nodes $i$ and $j$ in $\mathcal{G}$ and $\mathbf{A}_{i,j}=0$, otherwise. Let $\rm\Sigma^\ast$ denote the covariance matrix of $\mathbf{X}$, $\rm\Sigma^\ast = \mathbb{E}[\mathbf{XX}^{\top}]$.
\subsection{Differential Privacy}\label{DP_properties}
The notion of DP is defined based on the concept of graph neighboring databases.
\begin{definition}\label{DP_def}
(DP \cite{dwork2014algorithmic})
A graph analysis mechanism $\mathcal{A}$ obeys $(\varepsilon,\delta)$-DP, if for all neighboring graphs $\mathcal{G}$ and $\mathcal{G^\prime}$ and all subsets $O$ of $\mathcal{A}$'s range:
  \begin{equation*}
    \begin{split}
          \mathbb{P}[\mathcal{A}(\mathcal{G}) = O]
      \leq{\exp(\varepsilon)}\times{\mathbb{P}[\mathcal{A}(\mathcal{G^\prime}) = O] + \delta}.
    \end{split}
  \end{equation*}
\end{definition}
{\noindent}The parameter $\varepsilon$ denotes the privacy level, while a positive $\delta$ allows the algorithm to yield an unlikely output that reveals more information, yet only with probability bounded by $\delta$. When $\delta=0$, $\mathcal{A}$ is the called $\varepsilon$-DP. \\
\textbf{Gaussian Mechanism.} One way to satisfy DP is to add noise to the output of a query. In the gaussian mechanism \cite{dwork2014algorithmic}, in order to publish $f(\mathcal{G})$ where $f:\mathcal{G}\rightarrow\mathbb{R}^\ell$ while satisfying $(\varepsilon,\delta)$-DP, one publishes $\mathcal{A}(\mathcal{G})=f(\mathcal{G})+(Y_1,\cdots,Y_\ell)$, where $Y_i$ is drawn i.i.d. from Gaussian distribution $\mathcal{N}(0,S_2(f)^2\cdot\sigma^2)$ with
$\sigma=\sqrt{2\ln(1.25/\delta)}/\varepsilon$, where $S_2(f)=\mathop{\max}\limits_{\mathcal{G},\mathcal{G^\prime}}\|f(\mathcal{G})-f(\mathcal{G^\prime})\|_2$ is the global sensitivity of the quality function. \\
\hspace*{1em}In addition, DP also features the following main properties that are used in implementing PrivCom.
\begin{theorem}\label{DP_compos}
{\rm(Composition rule \cite{dwork2014algorithmic})} If $\mathcal{A}_1$ is $(\varepsilon_1,\delta_1)$-DP, and $\mathcal{A}_2$ is $(\varepsilon_2,\delta_2)$-DP, then $(\mathcal{A}_1\circ\mathcal{A}_2)$ is $(\varepsilon_1+\varepsilon_2,\delta_1+\delta_2)$-DP.
\end{theorem}
\begin{theorem}\label{Post_pro}
{\rm(Post-processing \cite{dwork2014algorithmic})}
If $\mathcal{A}$ is an $(\varepsilon,\delta)$-DP algorithm and $\mathcal{B}$ is an arbitrary data-independent mapping, then $\mathcal{B}\circ\mathcal{A}$ is also $(\varepsilon,\delta)$-DP algorithm.
\end{theorem}
\subsection{Katz Index}
\begin{definition}
($h$-order heuristics)
A method only involves the one-hop neighbors of two target nodes, which is called first-order heuristics.
Similarly, a method is calculated from up to two-hop neighborhood of the target nodes, which is called second-order heuristics.
Naturally, we can define $h$-order heuristics to be those heuristics which require knowing up to $h$-hop neighborhood of the target nodes.
\end{definition}
\begin{definition}\label{Close_subgra}
(Enclosing subgraph \cite{zhang2018link})
For a graph $\mathcal{G} = \left(\mathcal{V}, \mathcal{E}\right)$, given two nodes $x, y \in \mathcal{V}$, the $h$-order enclosing subgraph for $(x, y)$ is the subgraph $\mathcal{G}_{x, y}^{h}$ induced from $\mathcal{G}$ by the set of nodes $\{i \mid dis(i, x) \leq h$ or $dis(i, y) \leq h\}$, where $dis(i, x)$ denotes the shortest path distance between $i$ and $x$.
\end{definition}
\begin{definition}\label{gra_sig}
(Katz index \cite{katz1953new}) For a graph $\mathcal{G=(V,E)}$, given two nodes $x,y \in \mathcal{V}$, the Katz index for $\left(x,y\right)$ is defined as
$\mathbf{H}(x,y) = \sum_{l=1}^\infty \beta^l [\mathbf{A}^l]_{x,y}$,
where $l$ is the order, and $\beta$ is a decay factor that is smaller than the inverse of the spectral radius of $\mathbf{A}$.
\end{definition}
It is shown that high-order heuristics, such as Katz index, often performs better than
first-order and second-order ones \cite{lu2011link}.
It seems that, to learn high-order features, a larger $h$ will be needed, which is often infeasible on real-world social graphs due to time and space complexities.
For this, \cite{zhang2018link} presents an approximated Katz index that is actually capable of learning high-order graph features, even with a small $h$.
\begin{theorem}\label{Katz_AppThe}
(Katz index approximation \cite{zhang2018link})
Assume the damping factor $\beta < \frac{1}{d}$, where $d$ is the maximum node degree. The error between the approximated Katz index $\widetilde{\mathbf{H}}(x,y)=\sum_{l=1}^{2h+1}\beta^l \left[\mathbf{A}^l\right]_{i,j}$ and the real one is bounded as
$\left|\mathbf{H}(x,y) - \widetilde{\mathbf{H}}(x,y)\right|
      \leq \beta^{2h+2}\left(1-\beta d\right)^{-1} d^{2h+1}$.
\end{theorem}
\subsection{Oja's Method}
Let $\mathbf{v}_\tau\in\mathbb{R}^n$ denote the algorithm's estimate of the top eigenvector of $\rm\Sigma^\ast$ at $\tau$. Then, let $\eta$ denote learning rate, and $\mathbf{X}$ be a random sample. The update rule of Oja's algorithm \cite{jain2016streaming} is as follows:
\begin{equation}\label{Sim_oja}
  \begin{split}
    \mathbf{v}_\tau = \left(\mathbf{I} + \eta_\tau\mathbf{X}\mathbf{X}^\top\right)\mathbf{v}_{\tau-1}, \ \
    \mathbf{v}_\tau = \frac{\mathbf{v}_\tau}{\|\mathbf{v}_\tau\|_2}.
  \end{split}
\end{equation}
We see that Oja's method is a stochastic approximation algorithm that significantly reduces both space and time complexity. Note that the update in Eq. (\ref{Sim_oja}) computes only the top eigenvector. Yet, it can be generalized directly for the top-$k$ eigenvectors where $k>1$, by substituting $\mathbf{v}_\tau$ with matrix $\mathbf{V}_\tau \in \mathbb{R}^{k \times n}$, and by substituting the normalization step with column orthonormalization, for instance, by QR factorization.
\section{The Proposed Solution}\label{PrifreC_framwork}
\subsection{Overview of PrivCom}
The target of this paper is to release a sanitized graph $\widetilde{\mathcal{G}}$ that approximates the true community of the original graph $\mathcal{G}$ as closely as possible while satisfying the strict node-DP. Following a priori conclusion that the community structure is an important global pattern of nodes \cite{tu2019a}, we can move the burden of the utility guarantee on community structure to the global graph structure feature preservation. Motivated by this,
we provide an overview of our solution, called PrivCom, in Algorithm \ref{Priv_GraAlg}.
First, the approximated Katz matrix $\mathbf{\widetilde{H}}$ of $\mathbf{A}$ is calculated by Theorem \ref{Katz_AppThe} (line \ref{code:compute_approKatz}).
The fundamental advantage of adopting such a perspective is that we can learn global graph structure features from local subgraphs while greatly reducing the global sensitivity via a sensitivity regulation strategy (see section \ref{Sen_Analysis}).
Then, a private Oja algorithm is designed to approximate the eigen-decomposition, which yields the noisy Katz matrix via privately estimating eigenvectors and eigenvalues from extracted low-dimensional vectors (lines \ref{code:init_eigenvec}-\ref{code:gene_noisy_katz}). As mentioned before, with a fixed sensitivity, such approximation aims at tackling the impact excessive noise addition caused by Katz matrix requiring privacy budget splits.
Finally, based on the generated noisy Katz matrix, we reconstruct a synthetic graph (line \ref{code:gene_syn_graph}). \\
\hspace*{1em}In the following of this section, we discuss the key procedures
of PrivCom, namely, 1) sensitivity analysis; 2) private Oja algorithm; 3) synthetic graph generation.
\subsection{Sensitivity Analysis}\label{Sen_Analysis}
We now formally analyze the global sensitivity $S\left(\widetilde{\mathbf{H}}\left(i,j\right)\right)$. In this section, we will first obtain how the utility function $\widetilde{\mathbf{H}}\left(i,j\right)$ varies in neighboring databases. After that, we will formulate $S\left(\widetilde{\mathbf{H}}\left(i,j\right)\right)$ and demonstrate that, by tuning its decay factor, the noise scale enforced by node-DP can be greatly diluted. \\
\hspace*{1em}In this work, we leverage Katz index to capture the global structure feature of each node. From the definition of global sensitivity,
we have
\begin{definition}\label{Sen_Def}
(Global sensitivity)
  \begin{equation*}\label{Sen_Def_Eq}
     S\left(\widetilde{\mathbf{H}}\left(i,j\right)\right) = \max_{i,j \in \mathcal{V}, \mathcal{G}, \mathcal{G}^{\prime}}
     \left|\widetilde{\mathbf{H}}_{\mathcal{G}}\left(i,j\right)-\widetilde{\mathbf{H}}_{\mathcal{G}^\prime}^\prime\left(i,j\right)\right|,
  \end{equation*}
where $\mathcal{G}$ and $\mathcal{G}^\prime$ are neighboring graphs.
\end{definition}
\hspace*{1em}Intuitively, $S\left(\widetilde{\mathbf{H}}\left(i,j\right)\right)$ is the maximum change in the approximated Katz index in the output space if one node is missing. Consider the extreme case that the additional node connects to all other nodes, it is easy to observe that missing one node will most influence $|\mathcal{V}| - 1$ internal edges in a graph. Therefore, we have:
\begin{lemma}\label{Glo_Sen}
The global sensitivity of the Katz index for any $\left(x, y\right)$, $S\left(\widetilde{\mathbf{H}}\left(i,j\right)\right) = \left|\widetilde{\mathbf{H}}\left(i,j\right) - \widetilde{\mathbf{H}}^\prime\left(i,j\right)\right|$,
is enough to take 1, in which $0 < \beta < \frac{1}{n - 1}$.
  \begin{proof}\label{GloSen_Pro}
   We prove it by induction. When $l=1,\mathbf{A}_{i,j} \leq {d} \leq n - 1$ for any $(i,j)$ where $d$ represents the largest node degree in a graph. Therefore, the base case is correct. Now, assume by induction that $\left[\mathbf{A}^l\right]_{i,j}\leq{d^l}$ for any $(i,j)$, we have
     \begin{equation*}\label{GloSen_Pro_Equ_1}
       \begin{split}
          \left[\mathbf{A}^{l+1}\right]_{i,j}
          = \sum_{k=1}^n \left[\mathbf{A}^l\right]&_{i,k}
           \mathbf{A}_{k,j}
         \leq d^l\sum_{k=1}^n \mathbf{A}_{k,j}
         \leq d^ld
         = d^{l+1}.
            \end{split}
     \end{equation*}
     \begin{equation*}\label{GloSen_Pro_Equ_2}
       \begin{split}
         &\left|\widetilde{\mathbf{H}}\left(i,j\right) - \widetilde{\mathbf{H}}^\prime\left(i,j\right)\right|
          = \left|\sum_{l=1}^{2h+1}\beta^l\left[\mathbf{A}^l\right]_{i,j} - \sum_{l=1}^{2h+1}\beta^l\left[{\mathbf{A}^\prime}^l\right]_{i,j}\right| \\
         &= \left|\sum_{l=1}^{2h+1}\beta^l\left[\mathbf{A}^l - {\mathbf{A}^\prime}^l\right]_{i,j}\right|
         \leq \left|\sum_{l=1}^{2h+1}\beta^l d^l\right|
          \leq \left|\sum_{l=1}^{2h+1}\beta^l \left(n - 1\right)^l\right| \\
         &\stackrel{(1)}= \frac{1-\left[\beta \left(n - 1\right)\right]^{2h+1}}{\left[1 / \left( \beta \left(n - 1\right) \right)\right] - 1},
       \end{split}
     \end{equation*}
where $0 < \beta < \frac{1}{n - 1}$.
From step (1), we observe that with $\beta \in \left(0, \frac{1}{n-1}\right)$, its denominator is greater than 1, while its numerator is between 0 and 1. Naturally, the global sensitivity achieves the value in $(0, 1)$.
  \end{proof}
\end{lemma}
\begin{algorithm}[]
\caption{PrivCom Algorithm}\label{Priv_GraAlg}
\KwIn{graph matrix $\mathbf{A}$, privacy parameters $\varepsilon$ and $\delta$, learning rate $\eta$, an initial matrix $\mathbf{V} \in \mathbb{R}^{n \times k}$, total number of iterations $\gamma$, decay factor $\beta$.}
\KwOut{Synthetic graph $\mathcal{\widetilde{G}}$ with $(\varepsilon,\delta)$-DP.}
Compute the approximated Katz matrix $\mathbf{\widetilde{H}}$ of $\mathbf{A}$\;
\label{code:compute_approKatz}
Initialize $\widehat{\mathbf{V}}_0 = \mathbf{V}$ and let $\mathbf{X}\mathbf{X}^\top = \mathbf{\widetilde{H}}$\;
\label{code:init_eigenvec}
\For{$\tau = \mathrm{1},\ldots,\gamma$}
{
  $\widehat{\mathbf{V}}_\tau = \widehat{\mathbf{V}}_{\tau-1} + \eta\left(\mathbf{X}\mathbf{X}^\top \widehat{\mathbf{V}}_{\tau-1}+\mathcal{N}\left(0,\sigma^2\mathbf{I}\right)\right)$ \;
  \label{code:eigenvec_update}
  $\widehat{\mathbf{V}}_\tau = {\rm{QR}}(\widehat{\mathbf{V}}_\tau)$ \ $\triangleright$ the Gram-Schmidt decomposition that orthonormalizes the columns of $\widehat{\mathbf{V}}_\tau$\;
  \label{code:eigenvec_QR}
}
Derive $\widehat{\lambda}_\gamma = \sqrt{\|\mathbf{\widetilde{H}}\cdot\widehat{\mathbf{v}}_{\gamma}\|_2^2 + \mathcal{N}\left(0,\sigma^2\right)}$ for every $\widehat{\mathbf{v}}_{\gamma} \in \widehat{\mathbf{V}}_{\gamma}$\;
\label{code:noisy_eigenvalue}
Generate the noisy Katz matrix, $\mathbf{\widehat{H}}$, by $\widehat{\mathbf{V}}\widehat{\Lambda}\widehat{\mathbf{V}}^\top$ \ $\triangleright$ $\widehat{\Lambda} := \mathrm{diag}(\widehat{\lambda})$\;
\label{code:gene_noisy_katz}
Generate a synthetic graph by $\mathbf{\widehat{H}}$\;
\label{code:gene_syn_graph}
Return $\mathcal{\widetilde{G}}$\;
\end{algorithm}
\textbf{Sensitivity regulation strategy:} Note that in order to reduce the global sensitivity, the decaying factor $\beta$ is typically set to a small value. However, one difficulty is that whether the small setting will affect global graph structure feature capture. Towards this end, the following lemma reveals that the small $\beta$ only affects eigenvalues, which implies that with a proper $\beta$, we are able to reap a significant utility guarantee.
\begin{lemma}\label{Katz_EigenVec}
If $\left[\lambda,\mathbf{x} \right]$ is an eigen-pair of $\mathbf{A}$, then $\left[\mathcal{F}(\lambda),\mathbf{x} \right]$ is an eigen-pair of $\mathbf{\widetilde{H}}=\mathcal{F}(\mathbf{A})$.
  \begin{proof}
    According to the definition of an eigen-pair,
        $\mathbf{\widetilde{H}}\mathbf{x} = \lambda \mathbf{x}.$
    Then, we can easily get
        $\mathbf{\widetilde{H}}^2 \mathbf{x}
        = \mathbf{\widetilde{H}} \lambda \mathbf{x}
        = \lambda \mathbf{\widetilde{H}} \mathbf{x}
        = \lambda^2 \mathbf{x}.$
    By definition, $\left[\lambda^2,\mathbf{x }\right]$ is an eigen-pair of $\mathbf{A}^2$. By using the definition of $\mathbf{\widetilde{H}}=\mathcal{F}(\mathbf{A})$ and repeating the above process, we have
            $\mathbf{\widetilde{H}}\mathbf{x}
          = \mathcal{F}(\mathbf{A})\mathbf{x}
          = (\beta\lambda + \beta^2\lambda^2 + \ldots + \beta^{(2h+1)}\lambda^{(2h+1)})\mathbf{x}
          = \mathcal{F}(\lambda)\mathbf{x},$
    which completes the proof.
  \end{proof}
\end{lemma}
\subsection{Private Oja Algorithm}
\noindent\textbf{First-cut solutions:} To reap Katz matrix satisfying $(\varepsilon,\delta)$-DP, a natural but naive approach is to inject noise into each entry, that is $\mathbf{\widetilde{H}}(i,j) + \mathcal{N}(\cdot)$, in this matrix. More concretely,
the graph node number $n$ is fixed a priori, and the privacy budget is split across the Katz matrix, so $\varepsilon_1 = \cdots = \varepsilon_{n^2} = \frac{\varepsilon}{n^2}$.
Note that similar to the privacy budget $\varepsilon$, while the privacy parameter $\delta$ needs to be split, recall that the Gaussian mechanism, $\sigma=\sqrt{2\ln(1.25/\delta)}/\varepsilon$, whose logarithmic calculation enables us to ignore the impact caused by splitting $\delta$.
An alternative solution is to map this matrix to an $n^{2}$-dimensional vector, and then inject Gaussian noise, $\mathcal{N}\left(0,\sigma^2\mathbf{I}_{n^2 \times n^2}\right)$, into this vector. Unfortunately, due to the large scale of social graphs, it is unpractical to calculate this identity matrix $\mathbf{I}$ with size $n^2 \times n^2$. \\
\hspace*{1em}At a high level, we tackle this problem by leveraging a stochastic algorithm for eigen-decomposition approximation. In particular, we use Oja's algorithm \cite{AllenZhu2017FirstEC}, which computes top eigenvectors of a matrix with a stochastic access to the matrix itself. That is, if we want to compute the top eigenvector of $\mathbf{\widetilde{H}}$, we can use the following updates:
\begin{subequations}
  \begin{align}
    \widehat{\mathbf{V}}_\tau &= \widehat{\mathbf{V}}_{\tau-1} + \eta\mathbf{X}\mathbf{X}^\top\widehat{\mathbf{V}}_{\tau-1}, \
    \widehat{\mathbf{V}}_\tau = {\rm{QR}}(\widehat{\mathbf{V}}_\tau), \label{OjaAlo_Eq_a} \\
    \widehat{\mathbf{V}}_\tau &= \widehat{\mathbf{V}}_{\tau-1} + \eta\left(\mathbf{X}\mathbf{X}^\top+\mathcal{N}_\tau\right)\widehat{\mathbf{V}}_{\tau-1}, \label{OjaAlo_Eq_b} \\
    \widehat{\mathbf{V}}_\tau &= \widehat{\mathbf{V}}_{\tau-1} + \eta(\mathbf{X}\mathbf{X}^\top\widehat{\mathbf{V}}_{\tau-1} + g_\tau),\label{OjaAlo_Eq_c}
  \end{align}
\end{subequations}
where $\mathbf{XX}^\top = \mathbf{\widetilde{H}}$. For example, we can update $\widehat{\mathbf{V}}_\tau$ using Eq. (\ref{OjaAlo_Eq_b}) in which each entry of $\mathcal{N}_\tau$ is sampled i.i.d. from a Gaussian distribution calibrated to ensure DP. Even this algorithm in its current form does not mitigate privacy budget splits as we need to yield a high-dimensional matrix $\mathcal{N}_\tau$ with size $n \times n$ in each iteration. However, by observing that $\mathbf{X}\mathbf{X}^\top\widehat{\mathbf{V}}_{\tau-1}=g_\tau \sim \mathcal{N}\left(0,\sigma^2 \mathbf{I}_{n \times k}\right)$ (because the sensitivity of $g_\tau$ is 1, see Lemma \ref{Max_Sensitivity_1}), we can now replace the generation of $\mathcal{N}_\tau$ by the generation of a low-dimensional matrix $g_\tau$, thus significantly reducing the privacy budget split for the Katz feature matrix. Note that after reaping noisy eigenvectors, the eigenvalues can be derived by
$\widehat{\lambda}_\gamma = \sqrt{\left\|\mathbf{\widetilde{H}}\cdot\widehat{\mathbf{v}}_{\gamma}\right\|_2^2 + \mathcal{N}\left(0,\sigma^2\right)}$
(see line \ref{code:noisy_eigenvalue}). Similar to Lemma \ref{Max_Sensitivity_1}, with
$0 < \beta \leq \frac{1}{(n-1)(n^\frac{3}{2}k^\frac{1}{2} + 1)}$,
taking $S(\left\|\mathbf{\widetilde{H}}\cdot\widehat{\mathbf{v}}_{\gamma}\right\|_2^2) = 1$ is enough to achieve $(\varepsilon, \delta)$-node-DP.
\begin{lemma}\label{Max_Sensitivity_1}
(Sensitivity of $g_\tau$) Let $\mathbf{\widetilde{H}\widehat{V}} = g_\tau$. With $0 < \beta \leq \frac{1}{(n-1)(n^\frac{3}{2}k^\frac{1}{2} + 1)}$, the global sensitivity of $g_\tau$, $S(g_\tau)$, is 1.
  \begin{proof}\label{Max_Sensitivity_1_pro}
    To use Gaussian mechanism, we map the matrix $(\mathbf{\widetilde{H}\cdot\widehat{V}})_{n \times k}$ to vectors. Because for vectors, the value of $\|\cdot\|_2$ is equal to $\|\cdot\|_F$, so that the vectors and matrices can be uniformly quantified using $\|\cdot\|_F$. Thus, we have:
    \begin{equation*}
      \begin{split}
            \|\mathbf{\widetilde{H}\widehat{V}}\|_F
      \leq& \|\mathbf{\widetilde{H}}\|_F \|\mathbf{\widehat{V}}\|_F
      \leq  n^\frac{3}{2}k^\frac{1}{2} \max_{x,y}\left|\mathbf{\widetilde{H}}(x,y)\right| \max_{x,y}\left|\mathbf{\widehat{V}}(x,y)\right| \\
      \leq& \frac{ n^\frac{3}{2}k^\frac{1}{2} - n^\frac{3}{2}k^\frac{1}{2} \left[\beta \left(n - 1\right)\right]^{2h+1} }{ \left[1 / \left( \beta \left(n - 1\right) \right)\right] - 1 }
      \leq 1.
      \end{split}
    \end{equation*}
  \end{proof}
\end{lemma}
\begin{theorem}\label{Priv_Pro}
PrivCom satisfies $\Sigma_{i=1}^{\gamma + k}(\varepsilon_i,\delta_i)$-DP with regard to a change in a graph's node.
  \begin{proof}
    For each iteration, the low-dimensional projection vectors with Gaussian noise (see line \ref{code:eigenvec_update}) make the output eigenvector obey $(\varepsilon,\delta)$-DP. With composition rule (see Theorem \ref{DP_compos}), after $\gamma$ iterations, we easily get that the output eigenvector obeys $\Sigma_{i=1}^\gamma(\varepsilon_i,\delta_i)$-DP. Then, following this private eigenvector and the approximated Katz matrix that potentially leaks privacy, with Gaussian noise added, we enable the eigenvalues to be $\Sigma_{i=1}^k(\varepsilon_i,\delta_i)$-DP. According to the post-processing property of DP, we derive that PrivCom maintains $\Sigma_{i=1}^{\gamma + k}(\varepsilon_i,\delta_i)$-DP with regard to a change in a node of a graph.
  \end{proof}
\end{theorem}
\subsection{Synthetic Graph Generation}
In this section, we aim to reconstruct a synthetic graph from the noisy Katz matrix. We will need the following theorem.
\begin{theorem}\label{GraGene_The}
(Lemma 9.4.1. of \cite{Spielman2012TreeDis}).
Let $\mathbf{H}$ be a matrix with $\mathbf{H}(i, j) = \text{Katz}_{i, j}$ for all $i, j \in[n]$. The Laplacian $\mathbf{L}$ of the solution $\mathcal{G}$ is $-2 \cdot\left[\left(\mathbf{I}-\frac{\mathbf{J}}{n}\right) \mathbf{H}\left(\mathbf{I}-\frac{\mathbf{J}}{n}\right)\right]^{+}$, where $\mathbf{I}$ is the $n \times n$ identity matrix and $\mathbf{J}$ is the $n \times n$ all ones matrix.
\end{theorem}
\hspace*{1em}Recently, \cite{Hoskins2018InferringNF} has shown that an analogous result to Theorem \ref{GraGene_The} holds for graph recovery from all pairs Katz similarity scores.
However, this formula is quite unstable and generally yields an output which is far from a graph Laplacian even when $\mathbf{H}$ is corrupted by a small amount of noise. So we instead compute a regularized estimate, $\widetilde{\mathbf{L}} = -2 \cdot\left[\left(\mathbf{I}-\frac{\mathbf{J}}{n}\right) \mathbf{\widehat{H}} \left(\mathbf{I}-\frac{\mathbf{J}}{n}\right) + \alpha{\mathbf{I}}\right]^{+}$,
where $\alpha > 0$ can be chosen, e.g., by line search. Normally, $\widetilde{\mathbf{L}}$ will not be a valid graph Laplacian, and therefore we remove any negative edge weights to obtain $\widetilde{\mathcal{G}}$. \\
\hspace*{1em}It is worth mentioning, here, that completing graph generation by Theorem \ref{GraGene_The} will consume $O(n^3)$ time complexity, which is not feasible for large scale graphs. Towards this end, an alternative solution \cite{courrieu2005fast}, which is based on a full rank Cholesky factorization, is employed to effectively address this problem with efficiency and effectiveness.
\section{Experiment and Analysis}\label{Experiment}
\noindent\textbf{Datasets.}
We make use of three standard real datasets in our experiments.
1) polblogs (PB for short) \cite{Ackland2005MappingTU}: is a network of US political blogs with 1,224 nodes and 16,718 edges.
2) petster-hamster \footnote{PH is available at http://konect.uni-koblenz.de/networks/; CG is available at http://snap.stanford.edu/data/index.html} (PH for short): is a social network containing friendships and family links, which has 2,426 nodes and 16,631 edges.
3) CA-GrQc (CG for short): is a collaboration network with 5,242 nodes and 14,496 edges. \\
\textbf{Baselines.}
We compare ours with two private graph publishing algorithms, namely dK-2 graph model \cite{wang2013preserving} and HRG model \cite{xiao2014differentially} with edge DP, to show how good the utility of our method is, even under rigorous node DP requirements. More concretely, the dK-2 graph model \cite{wang2013preserving}, which preserves the dK-2 series in graph anonymization while achieving $(\varepsilon, \delta)$-edge DP (referred to as dK-2). The HRG model \cite{xiao2014differentially} stores a cluster of nodes in the same branch of the HRG tree, guaranteeing edge DP via sampling possible HRG structures in the model space by Markov chain Monte Carlo. Note, however, that \cite{xiao2014differentially} achieves $\varepsilon$-edge DP with the global sensitivity, and therefore it cannot be directly employed to compare with ours. For a fair comparison, we replace the Laplace mechanism applied in \cite{xiao2014differentially} with the Gaussian mechanism, but keep the global sensitivity and variance both mechanisms consistent. \\
\textbf{Parameter Settings.}
Throughout all the experiments, we vary the privacy budget $\varepsilon\in[0.1, 1.5]$ for PrivCom and HRG, and $\varepsilon \in [5, 20]$ for dK-2.
For PrivCom, when DP is used, $\delta$ is supposed to be $10^{-5}$, a value that is generally considered safe \cite{abadi2016deep} because it implies that the definition of DP is true with a probability of $99.999 \%$. As mentioned before, for a fair comparison, we keep $\delta = \frac{1.25}{\exp(1)}$ for HRG. The iteration number is set to $\gamma=\min \left\{\frac{1}{\beta}, \frac{\left\|\mathbf{\widetilde{H}}\right\|^{2}}{\sigma \sqrt{n}}\right\}$ and the learning rate is $\eta = \frac{1}{\gamma\cdot\sigma\sqrt{n}}$. Following \cite{wang2013preserving}, we keep $\delta = 0.01$ for dK-2. In addition, the regularized parameter is $\alpha = 5 \times 10^{-6}$. Due to the randomness of the algorithms, we run each algorithm 100 times and report the average results. \\
\textbf{Evaluation Metrics.}
For every pair $(\mathcal{G}, \mathcal{\widetilde{G}})$, where $\mathcal{G}$ is a real-life graph and $\mathcal{\widetilde{G}}$ is a synthetic graph sampled from a model learned from $\mathcal{G}$, we evaluate the extent to which $\mathcal{\widetilde{G}}$ preserves the community structure of $\mathcal{G}$ via the average $F_1$ score (Avg-$F_1$) \cite{rossetti2017tiles}.
In particular, given the detected communities, each community is matched with the most similar one in the ground-truth communities, and the $F_1$ score of the two matched sets $c_1$ and $c_2$ is defined as
  \begin{equation*}\label{NF1_score}
    \begin{split}
      F_{1}\left(c_{1}, c_{2}\right)=2 \frac{prec\left(c_{1}, c_{2}\right) \times recall\left(c_{1}, c_{2}\right)}{prec\left(c_{1}, c_{2}\right)+recall\left(c_{1}, c_{2}\right)},
    \end{split}
  \end{equation*}
where $prec\left(c_{1}, c_{2}\right)=\frac{\left|c_{1} \cap c_{2}\right|}{\left|c_{1}\right|}\in\left[0, 1\right]$ and $recall\left(c_{1}, c_{2}\right)=\frac{\left|c_{1} \cap c_{2}\right|}{\left|c_{2}\right|}\in\left[0, 1\right]$.
\subsection{Community Structure Preservation}
\begin{figure}[]
  \centering
  \includegraphics[width=3.5in, height=2.5in]{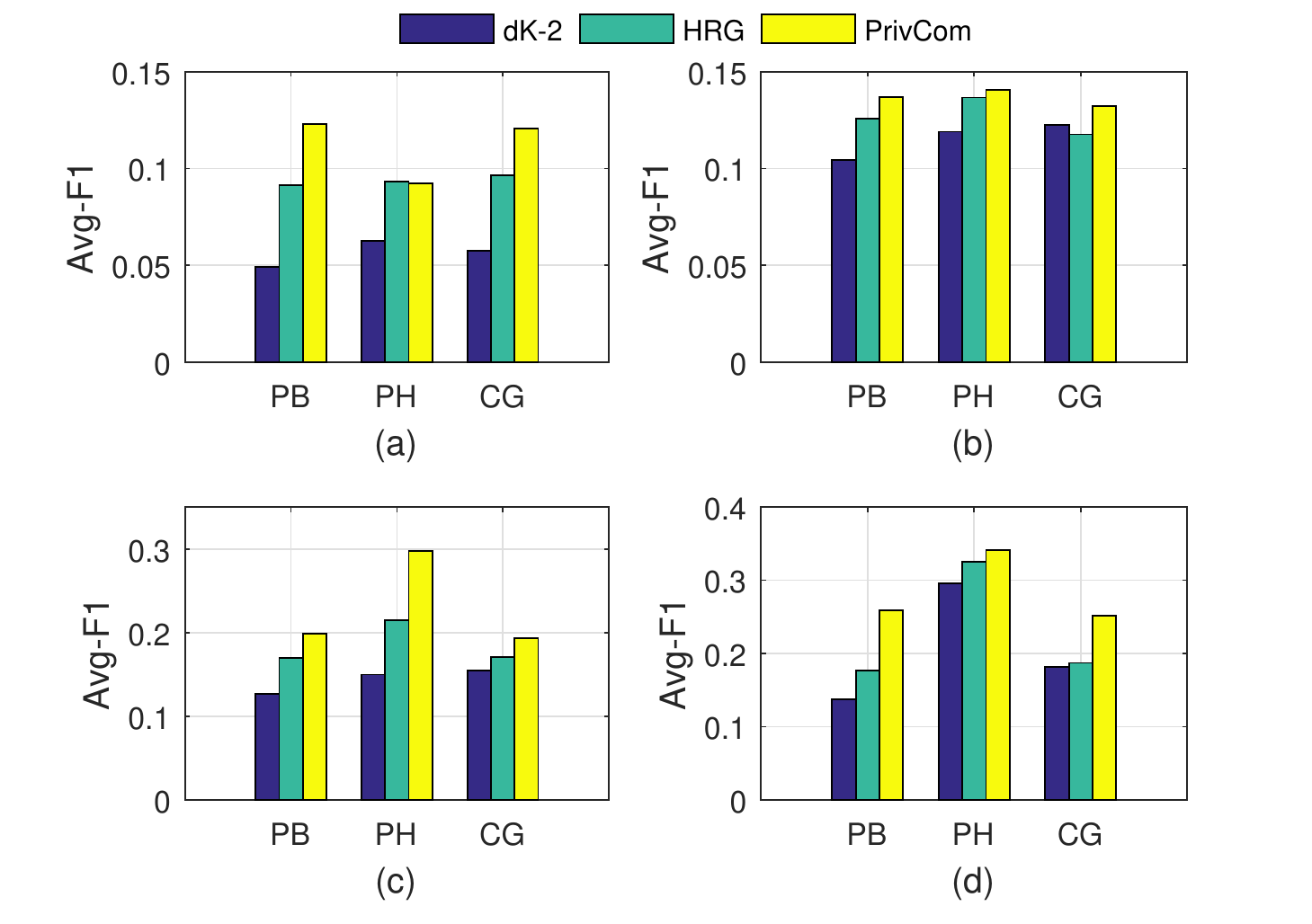}
  \caption{Comparison of Community Structure Preservation} \label{fig_four}
\end{figure}

Fig. \ref{fig_four} displays the comparison of the community structures found by Louvain \cite{blondel2008fast} in terms of the average $F_1$ score.
Here for HRG and PrivCom, we set $\varepsilon = 0.1, 0.5, 1, 1.5$ for Fig. \ref{fig_four}(a)-\ref{fig_four}(d), respectively. For dK-2, we set $\varepsilon = 5, 10, 15, 20$ for Fig. \ref{fig_four}(a)-\ref{fig_four}(d), respectively.
From Fig. \ref{fig_four}, we see that all algorithms perform in a similar way: the utility of the results is improved when the privacy budget $\varepsilon$ increases. This is, because, when $\varepsilon$ increases, a smaller amount of noise is required and a lower degree of privacy is guaranteed.
Besides, compared to HRG and PrivCom, we observe that even though dK-2 has larger privacy budgets, it still cannot reap competitive results. This reason is two-fold: i) this model only contains the linking information of two nodes, and contains little to global graph structure features; ii) a prohibitive amount of noise needs to be injected to mask the change of a single edge, which leads to poor overall data utility.
Another important observation from Fig. \ref{fig_four} is that in most cases, HRG reaps better performance than dK-2, while the performance gap between HRG and PrivCom is large on the three datasets tested. This phenomenon is in line with the analysis of section \ref{PrifreC_framwork}.
\section{Conclusion}\label{Conclusion}
In this paper, we have proposed PrivCom, a node-DP based graph publishing algorithm. The underlying highlights lie in the following aspects:
i) to reduce the huge sensitivity, we devise a Katz index-based private graph feature extraction method, which can capture global graph structure features while
greatly reducing the global sensitivity via a sensitivity regulation strategy; and ii) with a fixed sensitivity, the feature matrix generated by Katz index, requiring privacy budget splits, potentially mitigates global structural utility. For this, we design a private Oja algorithm approximating eigen-decomposition, which yields the noisy Katz matrix via privately estimating eigenvectors and eigenvalues from extracted low-dimensional vectors. Formal privacy analysis and experimental results on real datasets show that the released graph by PrivCom satisfies $(\varepsilon,\delta)$-node-DP while preserving high data utility on community structure.
\section*{Acknowledgment}
The work was supported by the National Natural Science Foundation of China under grant 61772131.

\bibliographystyle{IEEEtran}
\bibliography{myref}
\end{document}